\documentclass[useAMS,usenatbib,referee]{biom}

\usepackage{amsmath}
\usepackage{amssymb}
\usepackage{latexsym}
\usepackage{graphics, graphicx}
\usepackage{epsfig,epsf,rotating}
\usepackage{epsf}
\usepackage{theorem}
\usepackage{bm}
\usepackage{booktabs}

\usepackage{longtable}
\usepackage[caption=false]{subfig}
\usepackage{ulem}
\usepackage{url}

\usepackage[usenames,dvipsnames]{color}
\usepackage{color,soul}
\usepackage{xcolor,framed}
\definecolor{shadecolor}{HTML}{FFF200}
\usepackage{ifthen}
\newboolean{isclean}
\setboolean{isclean}{false}
\ifthenelse{\boolean{isclean}}{

}{}

\title{Bayesian Modeling of Multiple Structural Connectivity Networks During the Progression of Alzheimer's Disease}
\author{Christine B. Peterson$^1$, Nathan Osborne$^2$, Francesco C. Stingo$^3$, Pierrick Bourgeat$^4$,\\
 {\bf James D. Doecke$^4$ and Marina Vannucci$^{2,*}$}\email{marina@rice.edu}\\
$^1$Department of Biostatistics, MD Anderson Cancer Center, Houston, TX, USA\\
$^2$Department of Statistics, Rice University, Houston, TX, USA\\
$^3$Department of Statistics, Computer Science, Applications ``G. Parenti", University of Florence, Italy\\  
$^4$Australian eHealth Research Centre, CSIRO Health and Biosecurity, Herston, QLD, Australia
}

\begin{document}

\date{
{Accepted to \textit{Biometrics}}, January 2020}

\pagerange{1--20} \pubyear{2020}

\volume{000}
\artmonth{Accepted January}
\doi{000}


\label{firstpage}


\begin{abstract}
Alzheimer's disease is the most common neurodegenerative disease. The aim of this study is to infer structural changes in brain connectivity resulting from disease progression using cortical thickness measurements from a cohort of participants who were either healthy control, or with mild cognitive impairment, or Alzheimer's disease patients. For this purpose, we develop a novel approach for inference of multiple networks with related edge values across groups. Specifically, we infer a Gaussian graphical model for each group within a joint framework, where we rely on Bayesian hierarchical priors to link the precision matrix entries across groups. Our proposal differs from existing approaches in that it flexibly learns which groups have the most similar edge values, and accounts for the strength of connection (rather than only edge presence or absence) when sharing information across groups. Our results identify key alterations in structural connectivity which may reflect disruptions to the healthy brain, such as decreased connectivity within the 
occipital lobe with increasing disease severity.
We also illustrate the proposed method through simulations, where we demonstrate its performance in structure learning and precision matrix estimation with respect to alternative approaches.\\
\end{abstract}

\begin{keywords}
AIBL study; Alzheimer's disease; Bayesian inference; Gaussian graphical model; MRI data
\end{keywords}

\maketitle

\section{Introduction}
Dementia is a leading cause of death, disability, and health expenditure in the elderly, with  Alzheimer's disease (AD) accounting for the majority of cases. Much research in AD aims at understanding how the disease mechanisms affect the brain, in an effort to aid in the diagnosis and treatment of those with AD. Here we are interested in exploring the changes in structural connectivity for different brain regions through the progression of the disease.

Traditional approaches to structural neuroimaging studies have focused on investigating cortical thickness, volume, and the rate of tissue loss as specific neurodegenerative biomarkers that relate to changes in the aging brain. More recently, attention has been given to the estimation of networks that capture the connectivity between cortical regions of interest and to the changes in connectivity that result from the progression of the neurological disease. It is widely known that correlated regions of interest are more likely to be part of a network and that networks are related to specific cognitive functions \citep{Bloch2013}. 
During the progression of neurodegenerative disease, a person has a varying amount of cortical tissue loss, depending on their disease stage. As such, ``connections'' assessed throughout the disease trajectory represent coordinated changes in brain tissue, which are reflected in cortical thickness measures.

Statistical methods for network inference are a powerful tool to gain insight into the complex interactions that govern brain connectivity networks. When all samples are collected under similar conditions or reflect a single type of disease, methods such as the graphical lasso \citep{Friedman2008} 
or Bayesian graphical approaches \citep{WangBGLasso, Wang2012}  
can be applied to infer a sparse graph and thereby learn the underlying network. These  have been successfully used for the estimation of structural brain connectivity networks. 

In studies where samples are obtained for different groups or subtypes of a disease, like the Australian Imaging, Biomarkers and Lifestyle (AIBL) study of ageing described below, separate estimation for each subgroup 
reduces statistical power by ignoring potential similarities across groups, while applying standard graphical model inference approaches to the pooled data across conditions leads to spurious findings. Recently, estimation methods for multiple graphical models have been proposed in the statistical literature, including penalization-based approaches that encourage either common edge selection or precision matrix similarity 
\citep{Guo2011, Cai2015}. 
In particular, \cite{Danaher2014} developed convex penalization schemes designed to encourage similar edge values (the fused graphical lasso) or shared structure (the group graphical lasso). 
More recent proposals encourage network similarity in a more tailored manner, assuming that the networks for each sample group are related within a tree structure \citep{Oates2014, Pierson2015}, or, more generally, within an undirected weighted graph \citep{Saegusa2016, Ma2016}. These methods assume that the relationships across groups are either known a priori or learned via hierarchical clustering. More flexible approaches that employ a Bayesian framework to simultaneously learn the networks for each group and the extent to which these networks are similar have been proposed in \cite{Peterson2015} and \cite{Shaddox2016}. More specifically, \cite{Peterson2015} proposed representing the inclusion of edges using latent binary indicators, and the sharing of edges across groups was encouraged via a Markov random field prior linking the indicators.
\cite{Shaddox2016} improved upon \cite{Peterson2015} by replacing the $G$-Wishart prior on the precision matrix within each group with a mixture prior that is more amenable to efficient sampling. However, \cite{Shaddox2016} still addresses only the inclusion or exclusion of edges, without consideration of edge strength or direction.

For the analyses of this paper, we propose a Bayesian Gaussian graphical modeling approach which retains the advantages of the approaches by \cite{Peterson2015} and \cite{Shaddox2016} in flexibly learning cross-group similarities within a joint framework, but that accounts for the similarity of edge values across groups, rather than only the binary presence or absence of those edges. Our framework allows us to not only learn the precision matrices within each group, but also to characterize the extent of shared edge values across the groups. Empirically, we demonstrate that this key feature results in a more accurate inference of the precision matrices. Unlike related approaches in the frequentist framework \citep{Pierson2015, Saegusa2016}, which require a separate, ad-hoc step to learn the cross-group relationships, we can simultaneously learn both the within-group and cross-group relationships. Furthermore, even though penalization approaches are more scalable, they provide only point estimates of large networks, which are often unstable given limited sample sizes. Within our Bayesian approach, we can better quantify uncertainty in the estimates.

When applied to the data from the AIBL study, our method demonstrates that the majority of structural connections are preserved across all groups, but participants with AD have structural connectivity that is most unique compared to the other groups. In comparison to separate Bayesian estimation methods, the proposed method is able to identify a larger number of connections, reflecting the benefit of borrowing strength across groups. The fused graphical lasso, on the other hand, selects very dense graphs, that likely include a larger proportion of false positives edges, as also suggested by simulation studies \textcolor{black}{in our current work and in previous investigations \citep{Peterson2015, Shaddox2016}. This issue was noted by \cite{Danaher2014}, who recommended an AIC-based criterion, which we apply here, as the best objective method for parameter selection, but acknowledged that cross-validation, AIC, and BIC tend to favor models that are too large; the tendency to select overly dense graphs was also observed for standard graphical lasso \citep{Liu2010}.}

\subsection{The AIBL study}\label{sec:intro_aibl}
Here, we focus on cortical thickness measurements from participants in the AIBL cohort who were either HC (healthy control), MCI (mild cognitive impairment) or had AD (Alzheimer's disease). As a marker for neurodegeneration, cortical thickness is used to assess the atrophy of the cortical grey matter (GM) using MR images, and has been proposed as a more stable parameter for AD diagnosis than volume/density measures, because it is a more direct measure of GM atrophy \citep{Singh2006}. Investigation into GM atrophy allows the approximate measurement of neuronal loss, which is one of the underlying hallmarks of neurodegenerative diseases. Analyses using cortical thickness have been shown to successfully separate AD from MCI and healthy control \citep{Querbes2009}. Our aim is to examine how the progression of AD affects the structural networks of the brain.

The rest of the paper is organized as follows: 
In Section \ref{sec:model} we describe the proposed Bayesian joint graphical modeling approach and the posterior inference. We return to the case study in Section \ref{sec:case_study} and apply our method to estimate structural connectivity networks in subjects from cognitively normal to AD.
In Section \ref{sec:sim} we perform a simulation study and compare performance with alternative approaches. We conclude with a discussion in Section \ref{sec:disc}.

\section{Proposed model}\label{sec:model}

  
Let $K$ represent the number of sample groups (e.g., HC, MCI and AD) and let $\mathbf{X}_k$ be the $n_k \times p$ data matrix (e.g., cortical thickness on $p$ brain regions) for the $k$th group, with $k = 1, \ldots, K$.
We assume that the observed values within each group arise from a multivariate normal distribution, where
each row of $\mathbf{X}_k$ corresponds to an independent observation following the distribution $\mathcal{N}(\boldsymbol{\mu}_k, \mathbf{\Sigma}_k)$. Since we are interested in the covariance structure, rather than the means, we assume that the data are centered by group, so that $\boldsymbol{\mu}_k = \boldsymbol{0}_k$ for $k = 1, \ldots, K$. The group-specific covariance matrix $\mathbf{\Sigma}_k$ has inverse $\mathbf{\Sigma}_k^{-1} = \mathbf{\Omega}_k \equiv (\omega_{k, ij})$.
The multivariate normal distribution has the special property that $\omega_{ij} = 0$ if and only if variables $i$ and $j$ are conditionally independent given the remaining variables \citep{Dempster}. Non-zero entries in the precision matrix $ \mathbf{\Omega}_k$ therefore correspond to edges in the group-specific conditional dependence graph $G_k$, which can be represented as a symmetric binary matrix with elements $g_{k,ij} = 1$  if edge $(i,j)$ is included in graph $k$, and equal to zero otherwise.


In the Bayesian framework, inference of a graphical model is performed by tackling two interrelated sub-problems: selecting the model and learning the model parameters. Model selection is driven by identifying the graph structures $G_k$, while the precision matrices $\mathbf{\Omega}_k$ are the key model parameters. Unlike many of the existing Bayesian approaches for multiple undirected graphical models, which are based on prior distributions that link groups through the graph structures $G_k$, in this paper we propose a novel prior that links the groups through the parameters $\mathbf{\Omega}_k$, accounting for edge strength rather than only edge presence or absence. The specification of such a prior requires some care as all precision matrices are constrained to be positive semidefinite.

\subsection{Prior formulation}
Our goal is to construct a prior on the precision matrices $\mathbf{\Omega}_1 \ldots, \mathbf{\Omega}_K$ that enables inference of a graphical model for each group, encourages similar edge values when appropriate, and allows for computationally tractable posterior inference.
There have been a number of prior distributions proposed for the precision matrix $\mathbf{\Omega}$ in a Gaussian graphical model. Early approaches required restrictive assumptions on the graph structure (in particular, decomposibility) to allow tractable sampling \citep{Dawid1993, Giudici1999}. Later methods included shrinkage priors \citep{WangBGLasso}, which offered computational scalability but not graph selection, and conjugate priors with no restriction on the graph structure \citep{Wang2012}, 
which, due to limited computational scalability, could only be applied in the moderate $p$ setting \textcolor{black}{with less than 100 variables in a single network}. 

Here, we build on the stochastic search structure learning (SSSL) model of \cite{Wang2015}, which assumes a normal mixture prior on the off-diagonal entries of the precision matrix, enabling graph selection with no restrictions on the graph structure within a computationally efficient sampling framework.  To achieve this, we define a joint prior distribution on the precision matrices $\mathbf{\Omega}_1$, \ldots, $\mathbf{\Omega}_K$ that encourages similarity across groups in terms of the off-diagonal elements of the precision matrices. Specifically, we consider the continuous shrinkage prior \citep{WangBGLasso,Wang2015} 
for $K$ networks defined as
\begin{equation} \label{omega_prior}
p(\mathbf{\Omega}_1 \ldots, \mathbf{\Omega}_K|\{\mathbf{\Theta}_{ij}: i < j\}) \propto \prod_{i<j} \mathcal{N}_K\big( \boldsymbol{\omega}_{ij}'| \mathbf{0}, \mathbf{\Theta}_{ij} \big) \prod_i \prod_k \text{Exp}(\omega_{k,ii}|\lambda/2)\mathbf{1}_{\mathbf{\Omega}_1 \ldots, \mathbf{\Omega}_K \in M^+},
\end{equation} 
where $\boldsymbol{\omega}_{ij} = (\omega_{1,ij}, \ldots, \omega_{K,ij})$ is the vector of precision matrix entries corresponding to edge $(i, j)$ across the $K$ groups,  $\lambda > 0$ is a fixed hyperparameter, and $M^+$ denotes the space of $p \times p$ positive definite symmetric matrices. The first term in the joint prior specifies a multivariate normal prior with covariance matrix $\mathbf{\Theta}_{ij}$ on the vector of precision matrix entries $\boldsymbol{\omega}_{ij}$ corresponding to edge $(i,j)$ across groups. To define a prior on $\mathbf{\Theta}_{ij}$, we work with the decomposition 
$\mathbf{\Theta}_{ij} = \text{diag}(\boldsymbol{\nu}_{ij}) \cdot \mathbf{\Phi} \cdot \text{diag}(\boldsymbol{\nu}_{ij})$,
where $\boldsymbol{\nu}_{ij}$ is a $K \times 1$ vector of standard deviations specific to edge $(i,j)$, and $\mathbf{\Phi}$ is a $K \times K$ matrix shared across all $(i, j)$ pairs with 1s along the diagonal. To ensure that $\mathbf{\Theta}_{ij}$ is positive definite, the only requirements are that the standard deviations $\nu_{k,ij}$ must be positive and $\mathbf{\Phi}$ must be a valid correlation matrix.
Given these constraints, we can then define a mixture prior on the edge-specific elements of $\boldsymbol{\nu}_{ij}$  that enables the selection of edges in each graph, and a prior on the off-diagonal entries of $\mathbf{\Phi}$ that allows us to model the relatedness of edge values across the sample groups. Following \cite{Wang2015},
the standard deviations $\nu_{k,ij}$ are set to either a large or small value depending on whether edge $(i,j)$ is included in graph $k$, that is $\nu_{k,ij} = v_1$ if $g_{k,ij} = 1$, and $\nu_{k,ij} = v_0$ otherwise.
The hyperparameters $v_1 > 0 $ and $v_0 > 0$ are fixed to large and small values, respectively. Small values of $v_0$ will shrink the value of $\omega_{k,ij}$ for edges which are not included in the graph towards 0. This prior indirectly encourages the selection of similar graphs in related networks. Specifically, a small value of $\omega_{k,ij}$ will encourage small values of $\omega_{l,ij}$ for any other group $l$ and in turn the exclusion of edge $(i,j)$ in both groups $k$ and $l$. Similarly, a large value of $\omega_{k,ij}$ will encourage large values of $\omega_{l,ij}$ and the inclusion of edge $(i,j)$ in groups $k$ and $l$.
Networks $k$ and $l$ are considered related if the posterior distribution of the $(k,l)$ element of $\mathbf{\Phi}$ is concentrated on relatively larger values.

For the prior on the graphs $G_1$, $\ldots$, $G_K$, we assume an independent Bernoulli distribution
\begin{equation}
\label{bern_prior}
p(G_1, \ldots, G_K) \propto \prod_{k=1}^{K}\prod_{i < j} \big\{\pi^{g_{k,ij}} (1-\pi)^{1-g_{k,ij}}\big\}.
\end{equation}
This prior is analytically defined only up to a normalizing constant. As discussed in \cite{Wang2015}, the unknown normalizing constant of prior (\ref{omega_prior}) and prior (\ref{bern_prior}) are proportional and cancel out in the joint prior on $(\mathbf{\Omega}_k,G_k)$. Consequently, the parameter $\pi$ is not exactly the prior probability of edge inclusion; however, as shown by \cite{Wang2015} the effect of these unknown normalizing constants on the posterior inference is extremely mild, and the parameter $\pi$ can be easily calibrated to achieve a pre-specified level of sparsity.   

Recall that $\mathbf{\Phi}$ is a correlation matrix, and must therefore have all diagonal entries fixed to 1 and be positive definite. To specify the prior on $\mathbf{\Phi}$, we rely on the joint uniform prior: 
\begin{equation} \label{Phi_prior}
p(\mathbf{\Phi}) \propto 1 \cdot \mathbf{1}_{\mathbf{\Phi} \in \mathcal{R}^K},
\end{equation} where $\mathcal{R}^K$ denotes the space of valid $K \times K$ correlation matrices i.e.\ positive definite symmetric matrices $\mathbf{\Phi}$ such that $\phi_{jk} = 1$ for all $j = k$ and $|\phi_{jk}| < 1$ for all $j \neq k$. When $\mathbf{\Phi}  = \mathbf{I}$, the precision matrices for each group are independent, and the proposed model reduces to that of \cite{Wang2015} applied separately to each sample group.

Alternative priors could be defined on the precision matrices $\mathbf{\Omega}_1$, \ldots, $\mathbf{\Omega}_K$ that ensure the support to be constrained to the space of symmetric positive semidefinite matrices $M^+$. However, our proposed prior has the key advantage of computational tractability. In the next section we show how we can define a sampler that is automatically restricted to the targeted support $M^+$. In our model cross-group similarity is defined by $\mathbf{\Phi}$, which links the elements of the precision matrices, whereas previous approaches \citep{Peterson2015,Shaddox2016} encouraged similarity through a joint prior on the adjacency matrices $G_1,\ldots, G_K$.

\subsection{MCMC algorithm for posterior inference} \label{sec:mcmc_main}
We rely on Markov chain Monte Carlo (MCMC) to generate a sample from the joint posterior. At a high level, the sampling steps are as follows (see also Supplementary Material):
\begin{itemize}
\item \textbf{\textit{Step 1}}: For each sample group $k = 1, \ldots, K$,  we first update the precision matrix $\mathbf{\Omega}_k$ using a block Gibbs sampler with closed-form conditional distributions for each column, as in \cite{Wang2015}, and then update $G_k$ by drawing each edge from an independent Bernoulli.

\item \textbf{\textit{Step 2}}: We sample the entire correlation matrix $\mathbf{\Phi}$ at once using a Metropolis-within-Gibbs step following the parameter expansion method of \cite{Liu2006}.
\end{itemize}

After discarding the results from the burn-in period, we take the median model \citep{Barbieri2004} as the posterior selected value for the graph $G_k$ for each group. Specifically, we select edges $g_{k,ij}$ with marginal posterior probability of inclusion $\geq$ 0.5, as in \cite{Wang2015}. To obtain a posterior estimate of the precision matrix consistent with the selected graph, we resample $\mathbf{\Omega}_k$ conditional on the posterior estimate of $\mathbf{\Phi}$ and the selected value of $G_k$.

\section{Structural connectivity patterns in the AIBL cohort} \label{sec:case_study}

\subsection{Subjects and MRI data processing}
We have disease stage information and measurements of cortical thickness across 100 regions of interest in the brain from a total of 584 subjects. Here we focus on imaging data and cognitive assessments from the last follow up time point available. The subjects were divided into four groups: high performing HC (hpHC, n=143), HC (n=145), MCI (n=148), and AD (n=148). To obtain this classification, subjects were first evaluated by a clinician for current diagnosis and categorized as HC, MCI, or AD. HC subjects were further divided into hpHC and HC using eight different cognitive composite scores representing different cognitive domains. Magnetic resonance imaging (MRI) was performed on each subject, and the resulting images were parcellated into 100 regions of interest (ROIs).  Mean cortical thickness was computed in each ROI, and used in subsequent analysis. This gave us data on $p=100$ brain regions for the $K=4$ groups of subjects. Within each group, data were centered. Additional details on the cognitive scoring and MRI data processing, along with a list of ROIs grouped by lobe of the brain, are provided in the Supplementary Material.

\subsection{Application of the proposed method}
The application of our model requires the specification of a few hyperparameters. Here we provide details on the specification we used to obtain the results reported below and refer readers to the sensitivity analysis found in the Supplementary Material for more insights on parameter selection. In particular, priors (\ref{omega_prior}) and (\ref{bern_prior}) require the choice of the hyperparameters $\nu_0$, $\nu_1$, and $\pi$. These were set to $\nu_0 = 0.01$, $\nu_1 = 15$, and $\pi = \frac{2}{(100-1)}$. The parameters $\nu_0$ and $\nu_1$ were chosen so that the network structure results were sparse, while the selection of $\pi$ was based on the default setting recommended in \cite{Wang2015}. As a guideline, increasing $\nu_0$ while holding the ratio between $\nu_0$ and $\nu_1$ fixed will result in sparser graphs, as shown in the sensitivity analysis, which agrees with the sensitivity analysis provided in \cite{Wang2015}. Increasing the ratio between $\nu_0$ and $\nu_1$ while holding $\nu_0$ fixed will likewise increase the sparsity of the inferred graphs.

The results we report below were obtained by running two MCMC chains with 20,000 iterations, after a burn-in of 5,000 iterations. Posterior probabilities of inclusion (PPI) for each edge were compared for the two chains to check for convergence. A correlation of 0.997 was found between these two posterior samples.  We also used the Gelman and Rubin's convergence diagnostic \citep{GelmanRubin1992} to check for signs of non-convergence of the individual parameters of the estimated $\Phi$ matrix and the estimated precision matrices. Those statistics were all below 1.1, clearly indicating that the MCMC chains were run for a sufficient number of iterations. The results reported here were obtained by pooling together the outputs from the two chains to give a total of 20,000 MCMC samples. 
 
\subsection{Results} \label{sec:cs_results}

\begin{figure}
\minipage{0.9\textwidth}
  \includegraphics[width=\linewidth]{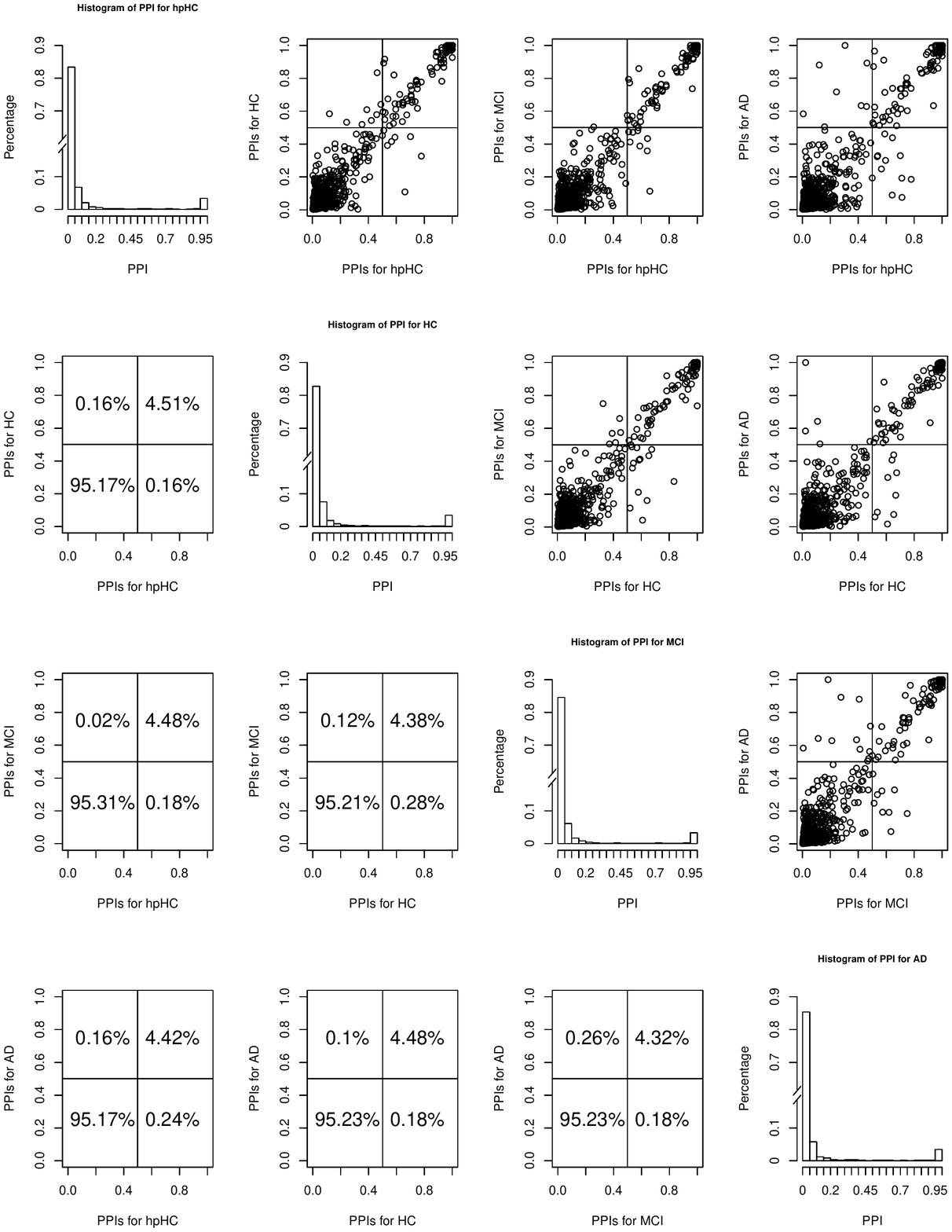}
\endminipage\hfill
\caption{Case study results discussed in Section \ref{sec:cs_results}. PPIs across the 4 groups of subjects. Plots on the diagonal show histograms of the PPIs for the individual groups. We introduced a break in the y-axis to allow better visualization of the small PPIs. Off-diagonal plots show scatter plots of the PPIs, on the upper triangle plots,  and percents of PPIs falling in each quadrant, in the lower triangle plots, for pairs of groups.}
\label{fig:PPIs}
\end{figure}

\begin{figure}
    \includegraphics[width=3.5in,page = 1]{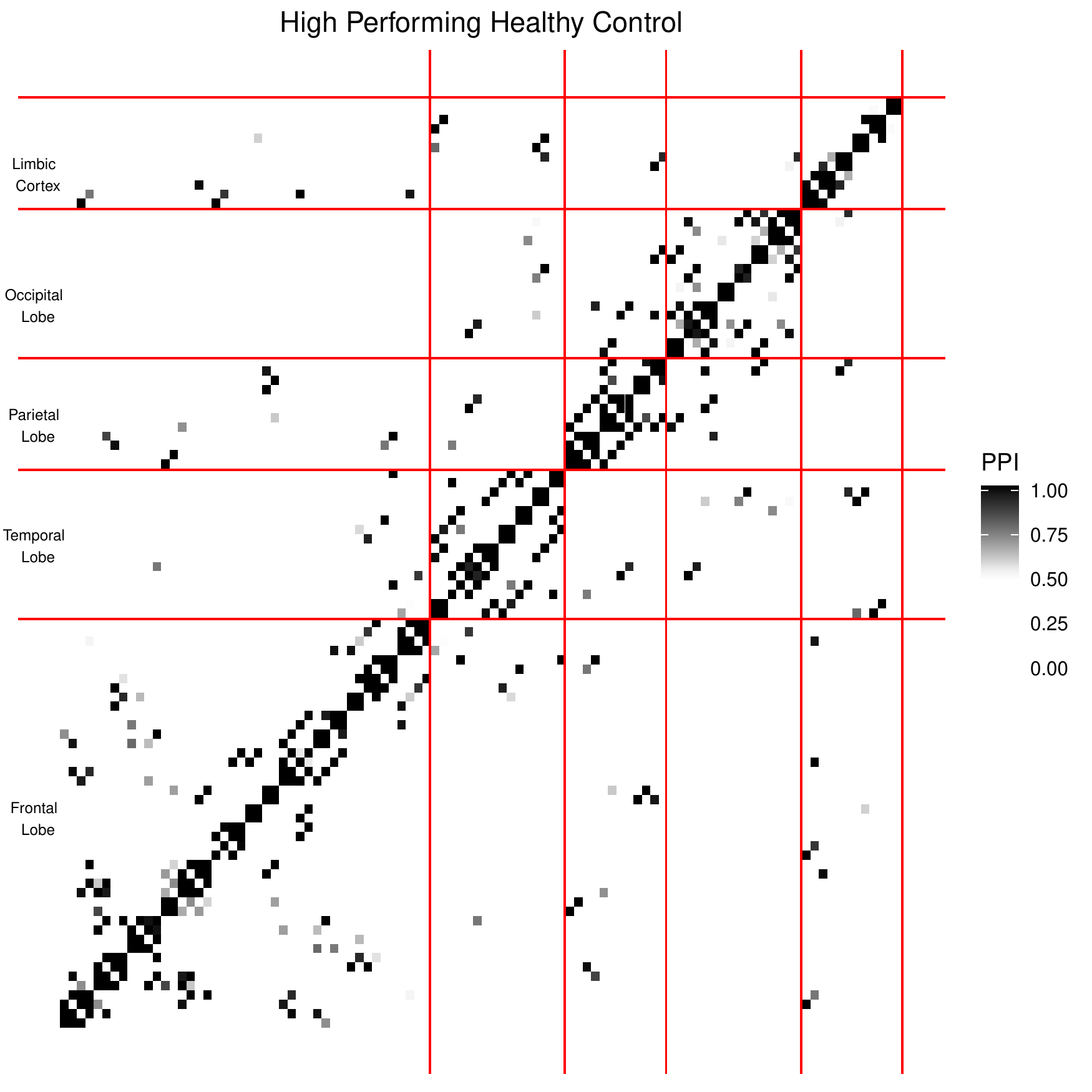}
    \includegraphics[width=3.5in,page = 2]{figure2.pdf}
    \includegraphics[width=3.5in,page = 3]{figure2.pdf}
    \includegraphics[width=3.5in,page = 4]{figure2.pdf}
    \caption{Case study results discussed in Section \ref{sec:cs_results}. Plot of the PPIs across the 4 groups of subjects. In each plot, ROIs are grouped within individual brain lobes.}
\label{fig:adjMatrices}
\end{figure}

Figure \ref{fig:PPIs} shows histograms of the PPIs for each group and scatter plots of the PPIs across pairs of groups. Off-diagonal plots show scatter plots of the PPIs, on the upper triangle plots,  and percents of PPIs falling in each quadrant, in the lower triangle plots, for pairs of groups.
In the scatter plots, the points in the upper right quadrants indicate edges that belong to the median model in both groups (shared edges), while points in the lower right and upper left quadrants indicate edges that were selected in one group but not the other (differential edges). The points in the lower left quadrant correspond to edges selected in neither group. These plots illustrate that the edge selection is fairly sparse overall, with a high concentration of PPIs close to 0 in the histograms, and that there are a number of edges which are strongly supported as shared across groups, as shown by the dense cluster of points in the upper right corner of the off-diagonal plots. 
Finally, we can see that many of the PPI values are the same across groups, as shown in the linear trend in the upper triangle plots. Although we generally do not observe a strong trend in terms of network differences across groups, we note that AD differentiates itself from the other groups most, because of the PPI values that vary (are relatively more dispersed from the linear trend) between AD and the other groups. Additionally, heatmaps of the PPIs within each group are shown in Figure \ref{fig:adjMatrices}. In these plots, the ROIs are groups within brain lobes, specifically, frontal, temporal, parietal, occipital and limbic cortex.
\textcolor{black}{These probabilities, which can only be obtained via a Bayesian approach, represent the confidence we have in the presence of each edge, and provide a useful summary of the uncertainty regarding edge selection.}
As expected, larger PPIs values are observed within lobes vs.\ across lobes for all disease stages. 

\begin{figure}
    \includegraphics[width=1.5in,page = 1]{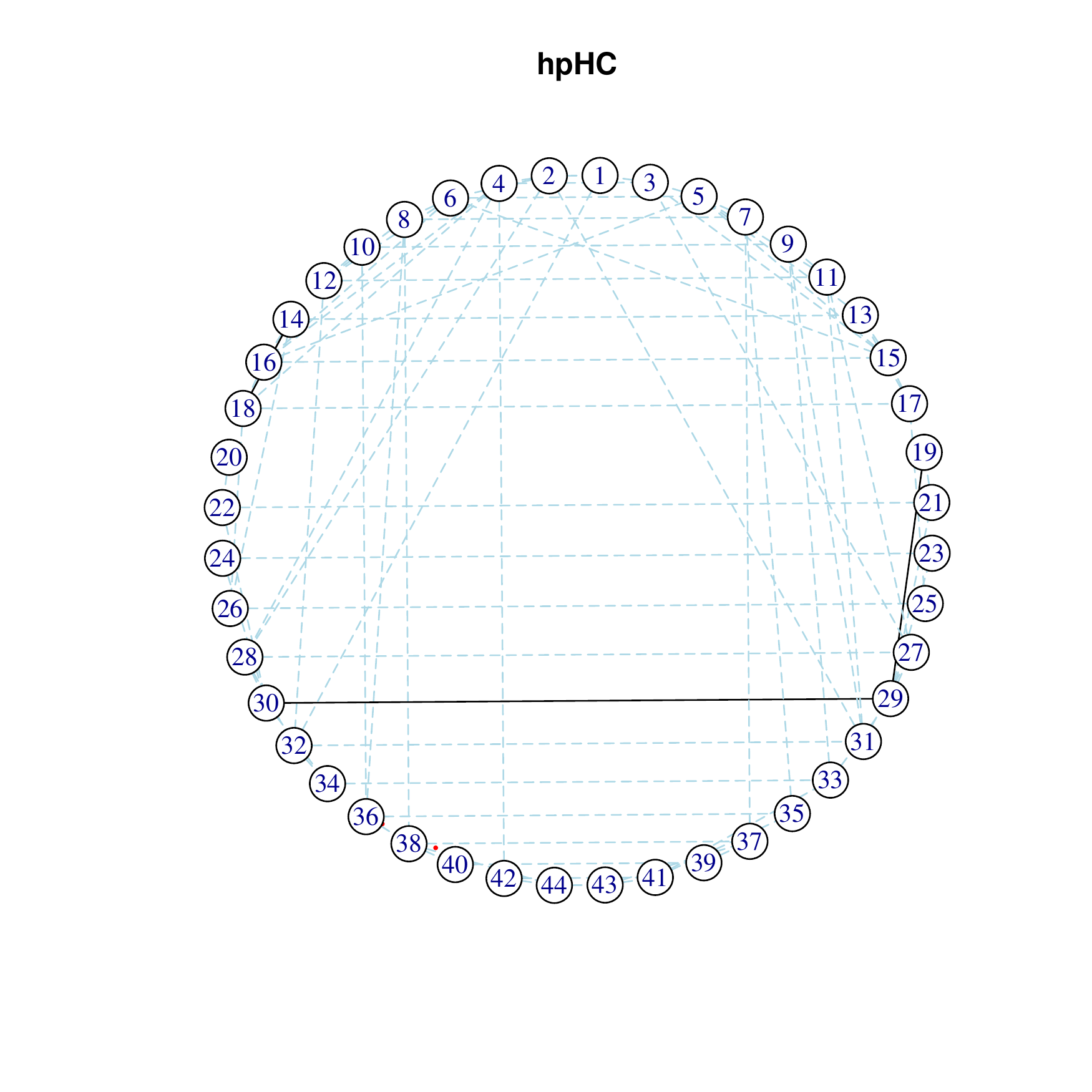}
    \includegraphics[width=1.5in,page = 2]{figure3.pdf}
    \includegraphics[width=1.5in,page = 3]{figure3.pdf}
    \includegraphics[width=1.5in,page = 4]{figure3.pdf}\\
    \includegraphics[width=1.5in,page = 5]{figure3.pdf}
    \includegraphics[width=1.5in,page = 6]{figure3.pdf}
    \includegraphics[width=1.5in,page = 7]{figure3.pdf}
    \includegraphics[width=1.5in,page = 8]{figure3.pdf}\\
    \includegraphics[width=1.5in,page = 9]{figure3.pdf}
    \includegraphics[width=1.5in,page = 10]{figure3.pdf}
    \includegraphics[width=1.5in,page = 11]{figure3.pdf}
    \includegraphics[width=1.5in,page = 12]{figure3.pdf} \\
    \includegraphics[width=1.5in,page = 13]{figure3.pdf}
    \includegraphics[width=1.5in,page = 14]{figure3.pdf}
    \includegraphics[width=1.5in,page = 15]{figure3.pdf}
    \includegraphics[width=1.5in,page = 16]{figure3.pdf}\\
    \includegraphics[width=1.5in,page = 17]{figure3.pdf}
    \includegraphics[width=1.5in,page = 18]{figure3.pdf}
    \includegraphics[width=1.5in,page = 19]{figure3.pdf}
    \includegraphics[width=1.5in,page = 20]{figure3.pdf}
   \caption{Case study results discussed in Section \ref{sec:cs_results}. Subnetworks corresponding to the frontal, temporal, parietal, occipital, and limbic lobes (from top to bottom), for the 4 groups of subjects, \textcolor{black}{where the edges shown are those selected in the median model. The left side of each circular array represents the left brain hemisphere, and the right side represents the right brain hemisphere. Blue lines indicate edges shared by all 4 groups, red lines indicate edges unique to an individual group, and black lines those shared by 2 or more groups.}}
\label{fig:igraph_lobes}
\end{figure}

To allow an in-depth view of the estimated networks, subnetworks corresponding to the individual lobes are shown in Figure \ref{fig:igraph_lobes}, where the edges shown are those selected in the median model; the estimated graphs $G_k$ for each group across all lobes are plotted in Supplementary Figure S2. In these circular plots, the left side represents the left brain hemisphere, and the right side represents the right brain hemisphere. In all plots, blue lines indicate edges shared by all 4 groups, red lines indicate edges unique to an individual group, and black lines those shared by 2 or more groups. 
The strongest pattern visible in the graphs are the horizontal blue lines connecting the corresponding regions in the right and left hemispheres of the brain.  The pattern of strong correlations between contralateral homologous regions of the cortex in structural imaging has been previously observed, for example by \cite{Mechelli2005}. 
 
Our findings are quantified in Table \ref{table:SharedSimilar}, which summarizes the numbers of edges included per group and shared across groups in the networks for all ROIs of Supplementary Figure S2 and the lobe-specific networks of Figure \ref{fig:igraph_lobes}. Within each subtable, the diagonal values represent the numbers of edges present in each group, and the off-diagonal values are the numbers of shared edges between pairs of groups.  Finally, the numbers of edges which are unique to a specific group is reported as values in parenthesis along the diagonals. 
From this, we see that the healthy control groups have slightly more edges than the cognitively impaired groups. We can also see that there is a decrease in connections in the occipital lobe as AD progresses. 
Additional ROI-specific patterns can be derived from Table S2 in Supplementary Material, which shows total number of edges for each ROI pair in each group. 

\begin{table}
\footnotesize
\begin{center}
\begin{tabular}{ |c| c c c c|  |c| c c c c|} 
\hline
All ROIs & hpHC & HC & MCI & AD &Frontal & hpHC & HC & MCI & AD \\
\hline
hpHC & 231 (1) &  &  &  &hpHC & 89 (1) &  &  &  \\ 
HC & 223 & 231 (3) &  &  &HC & 86 & 91 (2) &  & \\ 
MCI & 222 & 217 & 223 (1) &  &MCI & 87 & 85 & 87 (0) &  \\ 
AD & 219 & 222 & 214 & 227 (3) &AD & 86 & 89 & 85 & 89 (0) \\ 
\hline
\hline
Temporal & hpHC & HC & MCI & AD&Parietal & hpHC & HC & MCI & AD \\
\hline
hpHC & 25 (0) &  &  & &hpHC & 19 (0) &  &  &  \\ 
HC & 25 & 25 (0) &  &  &HC & 19 & 19 (0) &  &  \\ 
MCI & 25 & 25 & 25 (0) &  &MCI & 19 & 19 & 19 (0) &  \\ 
AD & 25 & 25 & 25 & 25 (0)  &AD & 19 & 19 & 19 & 19 (0) \\ 
\hline
\hline
Occipital & hpHC & HC & MCI & AD &Limbic & hpHC & HC & MCI & AD \\
\hline
hpHC & 30 (0) &  &  &  &hpHC & 12 (0) &  &  & \\ 
HC & 29 & 29 (0) &  &  &HC & 12 & 13 (0) &  &  \\ 
MCI & 27 & 26 & 27 (0) & & MCI & 11 & 11 & 11 (0) & \\ 
AD & 26 & 26 & 25 & 27 (1) &AD & 11 & 12 & 11 & 13 (1)  \\
\hline
\hline\\ \\
\end{tabular}
\end{center}
\caption{Case study results discussed in Section \ref{sec:cs_results}. Number of edges included per group and shared across groups in the networks for all ROIs of Supplementary Figure S2 and the lobe-specific networks of Figure \ref{fig:igraph_lobes}. Diagonal values represent the number of edges selected in each group, with values in parenthesis representing the number of edges that are unique to that group. Off-diagonal values indicate the numbers of shared edges between pairs of groups.}
\label{table:SharedSimilar}
\end{table}


Our method also produces estimated values of the elements of the $\mathbf{\Phi}$ matrix, which capture similarity in the precision matrix entries between the different subject groups. \textcolor{black}{Notably, as they are based on the joint posterior distribution, these values account for uncertainty in the estimation of the group-specific precision matrices.} 
\begin{equation*}
\begin{pmatrix}
\space & \text{hpHC} & \text{HC} & \text{MCI} & \text{AD} \\
\text{hpHC} & 1.000 &  &  &  \\ 
\text{HC} & 0.929 & 1.000 &  &  \\ 
\text{MCI} & 0.942 & 0.885 & 1.000 &  \\ 
\text{AD} & 0.865 & 0.940 & 0.883 & 1.000 \\ 
\end{pmatrix}
\end{equation*}
These values, which reflect the similarity in edge strength across groups,
provide a complementary look at the patterns of structural connectivity.
In particular, values of $\mathbf{\Phi}$ show that hpHC and AD are the least similar. They also show that HC and AD are related, which is supplemented by Table \ref{table:SharedSimilar} which shows that HC and AD have a large number of shared edges. The similarity of HC and AD may be caused by the way hpHC and HC were separated, as HC may have a higher propensity to develop AD. Our results also support similarity of the hpHC and MCI groups. Although these findings suggest there may be an underlying classification other than AD that influences the structural connectivity, the values we observe are generally large, supporting high degree of network similarity across groups. 

We conclude our analysis by summarizing the network structure of the estimated graphs via some graph metrics commonly used in neuroimaging \citep{Yao2010}. Specifically, we calculated the clustering coefficient $\gamma$, the absolute path length $\lambda$, and the small world coefficient $\sigma=\gamma/\lambda$. See \cite{Yao2010}, and references within, for a formal definition. 
\textcolor{black}{From a quantitative perspective}, if both $\lambda \approx 1$ and $\gamma>1$, and consequently $\sigma>1$, a network is said to exhibit small-world characteristics, \textcolor{black}{which means in a qualitative sense that any node can be reached from any other node in a small number of steps}.
Disconnected nodes were removed when calculating the characteristic path length. 
Based on the estimated values of $\lambda$ and $\gamma$, we obtain small world coefficients $\sigma$ of 1.717, 1.635, 1.627, and 1.475 for hpHC, HC, MCI, and AD, respectively. We observe that $\sigma$ is greater than 1 for all the groups, but steadily decreases during the progression of AD. Small-world characteristics in the brain network of AD have also observed by other authors \citep{He2008}. \textcolor{black}{Our conclusions on the differences in structural connectivity across groups are descriptive in nature, as our findings generally support a high degree of overlap in the structural connectivity networks.}

\subsection{Results from alternative approaches} \label{sec:cscomp}
For additional perspective, we compare our results to those of  the fused graphical lasso \citep{Danaher2014}, separate graph estimation in the Bayesian framework \citep{Wang2015}, and \textcolor{black}{the joint estimation approach of \cite{Shaddox2016}}.
For the fused graphical lasso, $\lambda_1$ and $\lambda_2$ were selected by performing a grid search to find the combination of values minimizing the AIC, as recommended in \cite{Danaher2014}. Separate Bayesian inference was applied with the same settings for $\nu_0$, $\nu_1$, $\lambda$, $\pi$ as in the linked method. \textcolor{black}{\cite{Shaddox2016} was applied with $\nu_0 = 0.50$,  $\nu_1 = 15$, $\lambda = 1$, $a = 1$, $b = 4$, $\alpha = 2$, $\beta = 5$, and $w = 0.5$.}

For each of the brain regions, Table \ref{table:casecomp} shows the number of total edges for each method on the diagonal, and the number of common edges on the off-diagonal.
Although the ground truth is not known, these results suggest that the proposed linked precision matrix method generally improves power over separate estimation: a large majority of the edges selected using separate estimation are also discovered under the proposed method, while separate estimation results in a slight increase in the number of edges across stages. 
We see a similarly large overlap of selected edges with the joint Bayesian estimation, though the joint Bayesian method leads to models that are more dense, due, in part, to the larger number of parameters of that model that control the sparsity.
The fused graphical lasso tends to select models which are even denser. This is because the AIC is not optimal for variable selection, tending to result in models which are not sufficiently sparse.
 
\begin{table}
\resizebox{1\textwidth}{!}{
\begin{tabular}{|c||c|c c c c|| c c c c|| c c c c|| c c c c|}
\hline
&& \multicolumn{4}{c}{\textbf{\color{black}{hpHC}}}& \multicolumn{4}{c}{\textbf{\color{black}{HC}}}& \multicolumn{4}{c}{\textbf{\color{black}{MCI}}}& \multicolumn{4}{c}{\textbf{\color{black}{AD}}}\\
\hline
 &  & Fused & Separate & Joint & Linked & Fused & Separate & Joint & Linked & Fused & Separate & Joint & Linked & Fused & Separate & Joint & Linked \\
\hline
All & Fused & 1486 &  &  &  & 1495 &  &  &  & 1345 &  &  &  & 1218 &  &  & \\
Regions & Separate & 167 & 168 &  &  & 175 & 175 &  &  & 181 & 181 &  &  & 185 & 185 &  & \\
 & Joint & 578 & 168 & 670 &  & 576 & 175 & 679 &  & 534 & 181 & 652 &  & 587 & 185 & 688 & \\
 & Linked & 229 & 142 & 215 & 231 & 229 & 147 & 218 & 231 & 222 & 160 & 221 & 223 & 226 & 165 & 223 & 227 \\
Frontal & Fused & 459 &  &  &  & 418 &  &  &  & 421 &  &  &  & 399 &  &  & \\
Lobe & Separate & 62 & 62 &  &  & 68 & 68 &  &  & 66 & 66 &  &  & 68 & 68 &  & \\
 & Joint & 204 & 62 & 216 &  & 198 & 68 & 220 &  & 196 & 66 & 212 &  & 211 & 68 & 225 & \\
 & Linked & 88 & 53 & 82 & 89 & 90 & 59 & 85 & 91 & 87 & 61 & 86 & 87 & 89 & 62 & 88 & 89 \\
Temporal & Fused & 73 &  &  &  & 76 &  &  &  & 66 &  &  &  & 65 &  &  & \\
Lobe & Separate & 23 & 23 &  &  & 20 & 20 &  &  & 24 & 24 &  &  & 24 & 24 &  & \\
 & Joint & 45 & 23 & 48 &  & 49 & 20 & 50 &  & 46 & 24 & 47 &  & 49 & 24 & 50 & \\
 & Linked & 25 & 22 & 24 & 25 & 25 & 19 & 25 & 25 & 25 & 22 & 25 & 25 & 25 & 23 & 25 & 25 \\
Parietal & Fused & 55 &  &  &  & 52 &  &  &  & 54 &  &  &  & 48 &  &  & \\
Lobe & Separate & 16 & 16 &  &  & 19 & 19 &  &  & 15 & 15 &  &  & 16 & 16 &  & \\
 & Joint & 40 & 16 & 40 &  & 40 & 19 & 40 &  & 37 & 15 & 37 &  & 34 & 16 & 35 & \\
 & Linked & 19 & 16 & 19 & 19 & 19 & 16 & 19 & 19 & 19 & 15 & 19 & 19 & 19 & 13 & 19 & 19 \\
Occipital & Fused & 97 &  &  &  & 104 &  &  &  & 89 &  &  &  & 75 &  &  & \\
Lobe & Separate & 22 & 22 &  &  & 27 & 27 &  &  & 25 & 25 &  &  & 23 & 23 &  & \\
 & Joint & 49 & 22 & 52 &  & 56 & 27 & 56 &  & 46 & 25 & 46 &  & 48 & 23 & 48 & \\
 & Linked & 30 & 22 & 29 & 30 & 29 & 23 & 29 & 29 & 27 & 23 & 27 & 27 & 27 & 22 & 27 & 27 \\
Limbic & Fused & 31 &  &  &  & 27 &  &  &  & 24 &  &  &  & 29 &  &  & \\
Lobe & Separate & 9 & 9 &  &  & 10 & 10 &  &  & 10 & 10 &  &  & 12 & 12 &  & \\
 & Joint & 14 & 9 & 16 &  & 16 & 10 & 16 &  & 13 & 10 & 13 &  & 17 & 12 & 17 & \\
 & Linked & 12 & 9 & 12 & 12 & 13 & 10 & 13 & 13 & 11 & 10 & 11 & 11 & 13 & 11 & 12 & 13 \\
\hline
\end{tabular} }
\caption{Comparison of case study results discussed in Section \ref{sec:cscomp}. For each group and brain region, diagonal values represent the total number of edges using the specified method, and off diagonal values represent the number of edges the two methods have in common.  \emph{Fused} is the fused graphical lasso of \protect \cite{Danaher2014}, \emph{Separate} is the separate Bayesian graph estimation with mixture priors of \protect \cite{Wang2015}, \emph{Joint} is the joint Bayesian estimation with mixture priors of \protect \cite{Shaddox2016}, and \emph{Linked} is the proposed approach.}
\label{table:casecomp}
\end{table}

\section{Simulation study} \label{sec:sim}
We present here a simulation study to compare performance across methods in learning graphs with related structure. The simulation is designed to mimic the real data application in terms of the number of variables, number of subjects per group, and graph structures.

We consider a setting with $K = 3$ groups, $p=100$ variables, and $n=150$ observations per group, where the underlying graph and precision matrix for each group are constructed as follows. $G_1$, the graph for the first group, consists of 5 communities, each with 20 variables. Within each community, the nodes are connected via a scale-free network. There are no connections across communities in $G_1$. The precision matrix entries in $\mathbf{\Omega}_1$ for edges are sampled independently from the uniform distribution on $[-0.6, -0.4]\cup[0.4, 0.6]$, while entries for missing edges are set to 0. To obtain $G_2$, five edges are removed from $G_1$ and five new edges added at random, so that now there are some cross-community connections. The entries in $\mathbf{\Omega}_2$ for the new edges are generated in a similar fashion as for $\mathbf{\Omega}_1$, while the entries for the edges removed are set to zero. To ensure positive definiteness,  $\mathbf{\Omega}_1$ and $\mathbf{\Omega}_2$ are each adjusted following the approach in \cite{Danaher2014}. To obtain $G_3$, 20 edges are removed from the graph for group 2, and the corresponding 20 entries in $\mathbf{\Omega}_2$ are set to zero to obtain $\mathbf{\Omega}_3$.
 These steps result in graphs $G_1$ and $G_2$ that share 180 of 185 edges (97.3\%), graphs $G_2$ and $G_3$ that share 165 of 185 edges (89.2\%), and graphs $G_1$ and $G_3$ that share 162 of the 185 edges in $G_1$ (87.6\%). The correlations between the off-diagonal elements of the precision matrices are 0.98 between $\mathbf{\Omega}_1$ and $\mathbf{\Omega}_2$, 0.94 between $\mathbf{\Omega}_2$ and $\mathbf{\Omega}_3$, and 0.93 between $\mathbf{\Omega}_1$ and $\mathbf{\Omega}_3$. To simulate the data, we generate $n$ samples per group from the multivariate normal $\mathcal{N}(0, \mathbf{\Omega}_k^{-1})$, for $k=1,2,3$. Below we report results obtained over 25 simulated data sets.

\subsection{Performance comparison}
We compare the following methods: fused graphical lasso \citep{Danaher2014}, group graphical lasso \citep{Danaher2014}, Bayesian inference applied separately for each group \citep{Wang2015}, Bayesian joint inference relating edge probabilities \citep{Shaddox2016}, and the proposed Bayesian joint inference method linking the precision matrix entries. For the lasso methods, the within-group penalty $\lambda_1$ and cross-group penalty $\lambda_2$ were selected using a grid search to identify the combination that minimize the AIC. Both separate Bayesian inference and the proposed linked precision matrix approach were applied using the parameter setting $\nu_0 = 0.01$, $\nu_1 = 0.1$, $\lambda = 1$, and $\pi = 2 / (p - 1)$. \cite{Shaddox2016} was applied using $\nu_0 = 0.05$,  $\nu_1 = 0.5$, $\lambda = 1$, $a = 1$, $b = 16$, $\alpha = 2$, $\beta = 5$, and $w = 0.5$, where the parameters were chosen to achieve a similar number of selected edges as obtained under the proposed linked precision matrix approach.

All Bayesian methods were run with 10,000 iterations as burn-in and 20,000 iterations for posterior inference. For the Bayesian methods, we take the posterior selected graph as the median model, and compute the posterior estimate of the precision matrices $\mathbf{\Omega}_k$ as the MCMC average when the precision matrices are resampled conditional on the graphs and the posterior estimate of $\mathbf{\Phi}$ from the initial run (for our method), or conditional on the graph using separate mixture priors (for separate and joint estimation approaches).

The performance across methods in terms of edge selection and differential edge selection is compared on the basis of true positive rate (TPR), false positive rate (FPR), Matthews correlation coefficient (MCC), and area under the curve (AUC). 
A detailed description of how these performance metrics were computed is provided in the Supplementary Material.
The performance results for graph and precision matrix learning are given in Table \ref{table_sim1}. In general, the Bayesian methods tend to favor sparser graphs, and achieve quite low false positive rates. The lasso methods tend to select somewhat denser graphs, and have correspondingly higher TPRs and FPRs. The proposed linked precision matrix method achieves the best overall performance, as demonstrated by its high MCC value. 
The AUC, which is computed across a range of model sizes, shows that the lasso methods and the proposed linked precision matrix approach have very good accuracy. For the lasso methods, the AUC was computed for multiple values of the cross-group penalty parameter while varying the within-group penalty, and the best was included here. Thus, the reported AUCs for these methods are likely to err on the optimistic side. Finally, the Frobenius loss is minimized under the proposed method.

\begin{table}
\begin{center}
\resizebox{1\textwidth}{!}{
\begin{tabular}{lll|l|l|l|l|l l l|l|l|l}
  \multicolumn{2}{c}{} &
  \multicolumn{6}{l}{\huge \textbf{All Edges}} & \multicolumn{5}{c}{\huge \textbf{Differential Edges}} \\
  \multicolumn{2}{c}{} 
  & \textbf{TPR} & \textbf{FPR} & \textbf{MCC} & \textbf{AUC} & \textbf{Fr Loss} & \textbf{\# edges} && \textbf{TPR} & \textbf{FPR} & \textbf{MCC} & \textbf{AUC} \\
  \cline{1-1}  \cline{3-8} \cline{10-13}
 Fused graphical lasso && 0.80  & 0.07  &  0.48 & \textbf{0.97}  & 0.065  &  461 &&  0.74  &  0.14   &  0.11  &  0.24  \\ 	 							
 &&  (0.01) &  (0.003) &  (0.01) &  (0.001) &  (0.001) & (15.1) &&  (0.01)  & (0.001)  & (0.003)  &  (0.01) \\     
  \cline{1-1}  \cline{3-8} \cline{10-13}
   Group graphical lasso &&  0.73 & 0.08 &  0.40 & 0.96 & 0.077 & 508 && 0.68  & 0.14 & 0.10 & 0.13  \\ 			
  &&  (0.01) &  (0.003) & (0.005) & (0.001) & (0.001) & (16.3) && (0.02)  & (0.004)  & (0.003) & (0.004) \\               
  \cline{1-1}  \cline{3-8} \cline{10-13}
   Separate estimation with  && 0.17 & 0.0002 & 0.40 & 0.89 & 0.099  & 31 && 0.16  & 0.01 & 0.10 & 0.84   \\     				
  \quad   mixture priors &&  (0.002) &  (3.0e-05) &  (0.003) &  (0.001)  & (0.001) & (0.5) && (0.01) & (2.0e-04)  & (0.01) & (0.01) \\            
  \cline{1-1}  \cline{3-8} \cline{10-13}
   Joint estimation with  && 0.57 & 0.03  & 0.47 & 0.89 &  0.327 & 236 &&  0.53 &  0.06 & 0.12  & 0.84  \\ 		     	     	     				
  \quad   mixture priors  && (0.004) & (3.0e-04) & (0.003) &  (0.002)  &  (0.003) & (1.6) && (0.02)  & (0.001) & (0.004) & (0.01)  \\          
  \cline{1-1}  \cline{3-8} \cline{10-13}
\textbf{Linked precision} &&  0.43 & 0.0002 &  \textbf{0.64} & 0.95  & \textbf{0.057}   & 77  && 0.22  &  0.003 &  \textbf{0.23} & \textbf{0.87} \\ 	     	     	     					
\quad    \textbf{matrix approach} &&  (0.01) &  (2.6e-05) &  (0.004) &  (0.001) &  (7.4e-04) & (1.1) &&  (0.01) &  (9.9e-05) &  (0.019) &  (0.01) \\    
\end{tabular}}
\end{center}
\normalsize
\caption{Performance summary across 25 simulated data sets. Comparison of true positive rate
  (TPR), false positive rate (FPR), Matthews correlation
  coefficient (MCC) and area under the ROC curve (AUC) for structure learning, and Frobenius loss (FL) for precision matrix estimation. The standard error of the mean is given in parentheses. \textcolor{black}{The methods compared are the fused and group graphical lasso of \protect \cite{Danaher2014}, separate Bayesian graph estimation with mixture priors of \protect \cite{Wang2015}, the joint Bayesian estimation with mixture priors of \protect \cite{Shaddox2016}, and the proposed linked precision matrix approach.}}
\label{table_sim1}
\end{table}

Based on the results in Table \ref{table_sim1}, the proposed method is conservative in the identification of differential edges, as indicated by its fairly low sensitivity and very high specificity. 
The proposed method achieves both the highest MCC and AUC across methods compared.
The high false positive rate of the lasso methods in selecting differential edges is partly due to the fact that they select a larger number of false positive edges overall, and may also reflect that they use a single penalty parameter to control cross-group similarity, which is not optimal when some groups have more similar dependence structure than others.

\noindent
Finally, the proposed linked precision matrix approach provides a posterior summary of cross-group similarity.
Specifically, the posterior estimated value of $\mathbf{\Phi}$ under the proposed linked precision matrix method is 
\begin{equation*}
\begin{pmatrix}
 1.0 & 0.65 &  0.63 \\
&  1.0 & 0.64 \\
& &  1.0 \end{pmatrix}. 
\end{equation*} Although the entries are fairly similar across groups, we can see that groups 1 and 2, which are the most similar to each other, have a higher value in the $\mathbf{\Phi}$ matrix.

Additional simulated scenarios with varying degrees of shared structure and edge values are included in the Supplementary Material. Results demonstrate that although the proposed method has the largest performance advantage when edge values across groups are in fact similar, it is robust to deviations from this setting, and performs similarly to separate Bayesian inference when there is no more overlap across groups than by random chance.


\section{Discussion} \label{sec:disc}
We have introduced a novel method for the joint analysis of multiple brain networks. The proposed approach allows flexible modeling of the cross-group relationships, resulting in relative measures of precision matrix similarity which fall in the $(0,1)$ interval. With respect to other methods for joint estimation, the proposed method not only shares information about the presence or absence of edges between groups, but also about the strength of those connections.
Building on the sampling framework of \cite{Wang2015} has allowed the proposed method to scale up \textcolor{black}{to around 100-150 variables}; the posterior sampling for a data set comprised of $p = 100$ ROIs and $K = 4$ groups took approximately 55 minutes for 1000 MCMC iterations in MATLAB on a laptop with a single Intel(R) Core(TM) i5-5200U CPU @ 2.20GHz and 16GB RAM. The proposed method was proven to be suitable for the analysis of multiple brain networks based on ROI measurements; in case interest is in larger networks, such as networks of voxels, more scalable approaches \textcolor{black}{focused on point estimation, such as lasso or EM algorithms \citep{Danaher2014, Li2019},} should be used.

We have applied our method to the analysis of structural data from the AIBL study on Alzheimer's disease, with the purpose of exploring the changes in structural connectivity for different brain regions through the progression of the disease. Our method has demonstrated that the majority of structural connections are preserved across all groups. 
Some of our findings are consistent with the literature on structural connectivity networks in Alzheimer patients: networks are fairly sparse and a number of edges are shared across groups. 
 
In theory, structural connectivity networks in Alzheimer's patients do not change dramatically with disease progression. Our findings confirm this theory, and support our assumption that all networks are similar to some extent, i.e.\ all elements of the $\mathbf{\Phi}$ matrix are non-zero. However, from a statistical modeling perspective, it might be of interest to replace the prior given in equation \eqref{Phi_prior} with a prior that assumes sparsity of the cross-group relationships. Such an extension is non-trivial due to the combination of constraints that $\mathbf{\Phi}$ must both be a positive-definite matrix and have all diagonal entries fixed to 1.

{\color{black}
\section*{Acknowledgements}
CBP, NO, and MV are partially supported by NSF/DMS 1811568/1811445. CBP is partially supported by NIH/NCI CCSG grant P30CA016672. 
}

\bibliographystyle{biom} 
\bibliography{refs}

{\color{black}
\section*{Supporting Information}
Web Appendices, Tables, and Figures referenced in Sections 3--5 are available with this paper at the Biometrics website on Wiley Online Library, along with Matlab scripts, R code and example data designed to resemble that of our real data application, also available online at {\url{https://github.com/cbpeterson/Linked_precision_matrices}}.  
}

\label{lastpage}
\end{document}